\documentclass[prd,reprint, nofootinbib,amsmath,amssymb, aps, floatfix]{revtex4-1}

\usepackage{subfigure}
\usepackage{amssymb,amsmath}
\usepackage{graphicx, xcolor, varwidth}
\usepackage{braket}
\usepackage{mathrsfs}
\usepackage{amsthm}
\newtheorem{thm}{Theorem}

\theoremstyle{remark}

\usepackage{bbold}
\usepackage{comment}

\begin{document}
\title{Restricted Quantum Focusing}
\author{Arvin Shahbazi-Moghaddam}
\affiliation{Stanford Institute for Theoretical Physics,\\ Stanford University, Stanford, CA 94305, USA}

\begin{abstract}
Quantum Focusing is a powerful conjecture, which plays a key role in the current proofs of many well-known quantum gravity theorems, including various consistency conditions, and causality constraints in AdS/CFT. I conjecture a (weaker) \emph{restricted} quantum focusing, which I argue is sufficient to derive all known essential implications of quantum focusing. Subject to a technical assumption, I prove this conjecture on brane-world semiclassical gravity theories which are holographically dual to Einstein gravity in a higher dimensional anti-de Sitter spacetime.
\end{abstract}

\maketitle

\section{Introduction}

Spacetime is emergent in quantum gravity: at length scales much larger than the Planck length, an approximate semiclassical description emerges where local quantum fields propagate on a smooth spacetime manifold.

Despite its approximate nature, semiclassical gravity quantifies and explains deep quantum gravity concepts in simple geometric terms. The generalized entropy is central to this story. Let $B$ be a partial Cauchy slice, such that $\partial B$ is a smooth codimension-two spacelike submanifold. The generalized entropy of $B$ is defined as~\cite{Bekenstein:1973ur, tHooft:1984kcu, Susskind:1994sm}
\begin{align}
    S_{\text{gen}}(B) = \frac{A(\partial B)}{4 G_d} + S(B)+\cdots
\end{align}
where $S(B)$ denotes the von Neumann entropy of the bulk fields in $B$, and the ellipsis denote sub-leading contributions to the generalized entropy from higher curvature corrections to Einstein gravity~\cite{Dong:2013qoa}. 

The importance of the generalized entropy becomes particularly evident in the context of AdS/CFT~\cite{Maldacena:1997re}. Any CFT subsystem is dual to a quantum extremal region $B$, i.e., a stationary point of the generalized entropy functional~\cite{Engelhardt:2014gca, Dong:2017aa}, with $S_{\text{gen}}(B)$ equal to the boundary subsystem's von Neumann entropy.\footnote{This is a special case of a more complicated story~\cite{Akers:2020pmf}. However, these complications can be ignored in a very large class of states.} This has, for example, led to a derivation of the Page curve~\cite{Page:1993wv, Penington:2019npb, Almheiri:2019psf}, extending even beyond AdS/CFT~\cite{Hartman:2020swn, Gautason:2020tmk}. Furthermore, quantum extremal regions dictate salient features of the holographic bulk-to-boundary map, resulting, for instance, in concrete proposals for its computational complexity~\cite{Brown:2019rox}. This has important consequences for the reconstruction of the black hole interior~\cite{Brown:2019rox, Engelhardt:2021qjs}, and has further sharpened some proposed resolutions to the firewall paradox in evaporating black holes~\cite{Almheiri:2012rt, Harlow:2013tf, Akers:2022qdl}. In addition, the generalized entropy outside black hole apparent horizons\footnote{More accurately, quantum minimar surfaces~\cite{Engelhardt:2017aux, Engelhardt:2018kcs, Bousso:2019dxk}.} has been identified with a coarse-grained entropy, giving the generalized second law of such horizons a statistical explanation~\cite{Engelhardt:2017aux, Engelhardt:2018kcs, Bousso:2019dxk}.

The quantum focusing conjecture (QFC)~\cite{Bousso:2015mna}, the quantum analogue of the classical focusing theorem, is a powerful constraint in semiclassical gravity, whose implications are at the heart of the above discoveries' consistency. For example, the QFC is a crucial assumption in various existence proofs of quantum extremal regions~\cite{Akers:2019lzs, Brown:2019rox, Engelhardt:2021mue}, and their compatibility with causality on the boundary CFT~\cite{Akers:2019lzs, Akers:2016aa}. The QFC also implies the quantum Bousso bound, quantum singularity theorems~\cite{C:2013uza, Bousso:2022tdb}, the generalized second law of causal horizons and holographic screens~\cite{Bousso:2015eda}, and the quantum null energy condition~\cite{Bousso:2015mna, Bousso:2016aa, Balakrishnan:2017bjg, Ceyhan:2018zfg}.

\begin{figure}
\includegraphics[width=.45\textwidth]{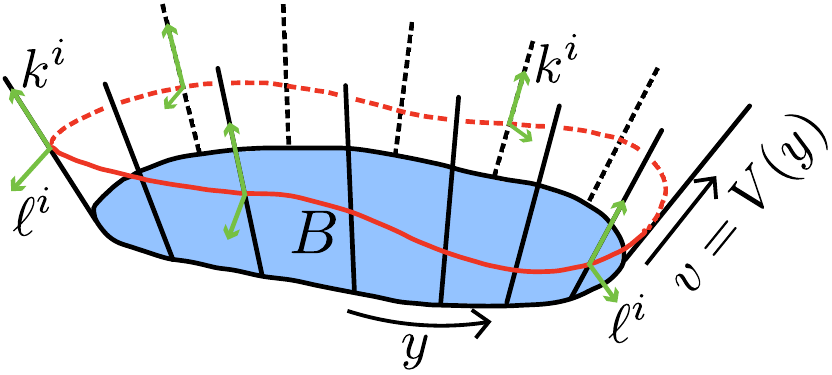}
\caption{Given a partial Cauchy slice $B$ (shown in blue), we define the null hypersurface $N^+(B)= (\dot{J}^+(B) - B) \cup \partial B$ whose generators are depicted with straight black lines with tangent vectors $k^i$. A future Cauchy slice $\Sigma_V$ intersects $N^+(B)$ at $v=V(y)$, depicted in red. On any surface $v=V(y)$, we can define a null vector field $\ell^i$ orthogonal to it, which is outward and past-directed. The quantum expansions $\Theta_{(k)}$ and $\Theta_{(\ell)}$ of $B_V$ are given by the rates of change of $S_{\text{gen}}(B_V)$, per unit transverse area, per unit affine length, as the region is deformed locally at $\partial B_V$ along the $k^i$ and $\ell^i$ directions respectively.}\label{fig-JB}
\end{figure}

Despite its prominent role in semiclassical gravity and holography, quantum focusing remains without a general proof. The goal of this paper is to (partly) fill this gap. In Sec.~\ref{sec:rqfc1}, we conjecture a condition weaker than the QFC, which is sufficient to replace it in the aforementioned applications. Sec.~\ref{sec:braneproof} includes a proof of this and another relevant constraint, on brane-world semiclassical gravity theories that are holographically dual to Einstein gravity in an asymptotically (locally) Anti-de Sitter spacetime (henceforth, referred to as brane-world gravity). We conclude in Sec.~\ref{sec:disc} with a discussion of some related ideas and future directions.

\section{The restricted quantum focusing conjecture}\label{sec:rqfc1}

We begin by defining some relevant objects. Let $J^+(B)$ be the causal future of $B$ and $N^+(B)= (\dot{J}^+(B) - B) \cup \partial B$ a null hypersurface with affine generators $k^i$ ($=\partial_v$). Now, let $\Sigma_{V}$ be a Cauchy slice nowhere to the past of $\partial B$ which intersects $N^+(B)$ at $v = V(y)\geq 0$ ($v=0$ on $\partial B$, and $y$ denote transverse coordinates on $\partial B$ which label the generators) and let $B_V =\Sigma_{V} \cap J^+(B)$ (See Fig.~\ref{fig-JB}).

Let $V_\lambda(y)$ be a one-parameter family of non-negative functions which satisfy $\partial_\lambda V_\lambda (y)\geq0$. The QFC states that~\cite{Bousso:2015eda}
\begin{align}\label{eq-QFC}
    \partial_\lambda \Theta_{(k)} (V_\lambda;y) \leq 0 \text{ for all $y$},
\end{align}
where
\begin{align}\label{eq-qexp}
\Theta_{(k)} (V;y) = \frac{4G_d}{\sqrt{h_V}} \frac{\delta S_{\text{gen}}(B_V)}{\delta V(y)},
\end{align}
is called the quantum expansion~\cite{Bousso:2015eda} of $B_V$ at $y \in \partial B_V$, and $h_V$ is the determinant of the induced metric on $\partial B_V$.

It is easy to see that the QFC implies:
\begin{align}
    \Theta_{(k)}(0;y)\leq 0 \text{, for } y \in \Gamma \subseteq \partial B \stackrel{V\rvert_{\partial B -\Gamma}=0}{\implies} \Theta_{(k)}(V;y)\leq 0,\label{eq-negexpansion}
\end{align}
where $V\rvert_{\partial B -\Gamma}=0$ means that $V$ is zero on all generators emanating from $\partial B -\Gamma$.

Interestingly, Eq. \eqref{eq-negexpansion} is all that is required of the QFC in the applications mentioned in the introduction. A look-alike condition, unrelated to the QFC, which is also crucial to the aforementioned applications, is\footnote{This condition involves a variation of the von Neumann entropy of $B$ under null deformations of $\partial B$ at \emph{different} points. This can be rewritten as an expression involving the von Neumann entropy of three subsystems, which by the strong sub-additivity of the von Neumann entropy acquires a sign~\cite{Wall:2009wi, Bousso:2015eda}. But Eq.\ref{eq-negexpansion2} also involves a contact term contribution from the Dong entropy piece of the generalized entropy~\cite{Dong:2013qoa}.}
\begin{align}
\Theta_{(\ell)} (0;y)\leq0 \stackrel{V(y)=0}{\implies} \Theta_{(\ell)} (V;y)\leq 0\label{eq-negexpansion2},
\end{align}
where for any region $B_V$, we define $\ell^i$ as the past outward directed vector field orthogonal to $\partial B_V$. Then, $\Theta_{(\ell)} (V;y)$ is defined in obvious analogy with Eq.\eqref{eq-qexp} (See Fig.~\ref{fig-JB}).

Throughout our discussion so far, we can interchange $J^+(B)$ with $D^+(B)$ (future domain of dependence of $B$). That is, we can consider inward deformations of $B$ along future-directed null geodesics orthogonal to $\partial B$. Then $k^i$ would be future-inward directed and $\ell^i$ past-inward directed. Together, conditions \eqref{eq-negexpansion} and \eqref{eq-negexpansion2}, along with their inward, and also time-reversed versions, imply all of the applications mentioned in (paragraph 4 of) the introduction.\footnote{To show this is a straightforward exercise in most cases which we leave to the interested and/or skeptical reader. Technically, an additional often-overlooked (and independent of the QFC) assumption is involved: that the loss of generators along $N^+$, which happens generically due to caustics and self-intersections, cannot increase the value of $S_{\text{gen}}$. Separately, To arrive at the quantum null energy condition, one needs to approach a classically stationary point $y$ on $\partial B$, through a family of surfaces which satisfy $\Theta_{(k)}=0$ at $y$ in the $G\to 0$ limit.} It is therefore highly desirable to prove them.

Here, I conjecture a \emph{restricted} quantum focusing condition, which states that
\begin{align}\label{eq-weakQFC}
    \Theta_{(k)}(V_{\lambda};y) = 0 \implies \partial_\lambda \Theta_{(k)}(V_\lambda;y) \leq 0.
\end{align}
Even though the restricted QFC is weaker than the QFC, it is sufficient to derive \eqref{eq-negexpansion}. To see this, pick any $V_{\lambda}$ such that $V_0(y)=0$ and $V_1(y)=V(y)$, which further satisfies the property that for each $y$, $\Theta_{(k)}(V_\lambda;y)$  is a differentiable function of $\lambda$. We expect that all physical states allow such a choice.\footnote{The reader might object that, for example, in shockwave geometries with a delta function energy sources, this is not the case. However, such delta function divergences only make sense as a distribution, and a physically reasonable state needs to involve a proper smearing of such delta functions which would then allow a differentiable choice.} Then, a violation of \eqref{eq-negexpansion} at some transverse point $y$ implies that there exists a $\lambda$ such that $\Theta_{(k)}(V_{\lambda},y)=0$, but $\partial_\lambda \Theta_{(k)}(V_\lambda,y) > 0$. Therefore, \eqref{eq-weakQFC} implies \eqref{eq-negexpansion}.

Similarly, the following inequality implies \eqref{eq-negexpansion2}:
\begin{align}\label{eq-weakQFCell}
    \Theta_{(\ell)}(V_{\lambda};y) = 0 \stackrel{\partial_\lambda V_\lambda(y)=0}{\implies} \partial_\lambda \Theta_{(\ell)}(V_\lambda;y) \leq 0
\end{align}

The rest of the paper is mainly devoted to proving conditions \eqref{eq-weakQFC} and \eqref{eq-weakQFCell} in brane-world gravity.

\section{A Proof of Restricted Quantum Focusing on brane-world gravity}\label{sec:braneproof}

We will introduce the brane-world setup briefly in subsection~\ref{sec3.1}, reviewing the salient points of the construction for our purposes, before delving into the proofs of conditions \eqref{eq-weakQFC} and \eqref{eq-weakQFCell} in subsection \ref{sec:proof}. For much more elaborate discussions of brane-world holography setups, see~\cite{Randall:1999vf, Emparan:2006ni, Emparan:1999wa, Verlinde:1999fy, Gubser:1999vj, Myers:2013lva, Chen:2020uac, Chen:2020hmv, Bousso:2020kmy, Karch:2000ct, Takayanagi:2011zk, Fujita:2011fp}.

\subsection{The brane setup}\label{sec3.1}
In the standard AdS/CFT setup, to compute the CFT$_d$ partition function holographically, one considers a cutoff surface at $z=\epsilon$, where $z$ is Fefferman-Graham (FG) radial coordinate of AdS$_{d+1}$, computes the bulk action including the Gibbons-Hawking-York terms. Then, as $\epsilon$ is sent to zero appropriate counter-terms are added to cancel divergences.

One way to think about brane-world holography is to instead consider a (physical) cutoff surface at a finite distance, with a metric that is free to fluctuate. The previously divergent contributions (no longer removed by counter-terms), may now be interpreted as induced gravity on the brane (e.g., as in the Randall–Sundrum model~\cite{Randall:1999vf}) which is coupled to a strongly-interacting holographic CFT. This is a semiclassical gravity theory on the brane which is holographically dual to a higher dimensional classical (Einstein gravity) theory.

Explicitly, consider the bulk action,
\begin{align}
    I_{\text{total}} = \frac{1}{16 \pi G_{d+1}}&\int d^{d+1}x \sqrt{-\bar{g}}~\left(\bar{R} + \frac{d(d-1)}{L^2}\right) \nonumber\\
   &+\frac{1}{8 \pi G_{d+1}}\int_{\text{brane}} d^dx\sqrt{-g}~ \left(K -T\right),
\end{align}
where $G_{d+1}$ and $L$ denote the bulk Newton's constant and the AdS radius respectively, $\bar{R}$ denotes the bulk Ricci scalar, $K = g^{ij}K_{ij}$ is the trace of the brane extrinsic curvature tensor $K_{ij}$. There also exists a brane tension term (proportional to $T$) which can fine-tune the brane location. The intrinsic metric on the brane is free to fluctuate, resulting in the equations of motion
\begin{align}\label{eq-braneeom123}
K_{ij} = (K-T) g_{ij}.
\end{align}

A practical way to find explicit brane solutions like this is to start with an asymptotically locally AdS$_{d+1}$ solution in FG coordinates,
\begin{align}
	ds^2 = \frac{L^2}{z^2}(dz^2 + \tilde{g}_{ij}(\tilde{x},z) d\tilde{x}^i d\tilde{x}^j),
\end{align}
with the condition (always achievable by an appropriate rescaling of $z$) that the smallest lengthscale on $\tilde{g}_{ij}(\tilde{x},z=0)$, denoted by $L_0$, satisfies $L_0 \gg L$.  Then the FG expansion remains valid at $z\sim L$, and furthermore, it is easy to check that, with $T =(d-1)/L$, the brane will be located at:
\begin{align}\label{eq-smallz}
z = L + O(L/L_0)
\end{align}
where the subleading correction are $\tilde{x}$ dependent in general. Importantly, using the FG expansion one can check that the induced gravity on the brane is Einstein gravity plus higher curvature corrections that are suppressed by powers of $L$~\cite{Myers:2013lva}:
\begin{align}\label{eq-branederivexp}
I_{\text{brane}} = \frac{1}{16 \pi G_d}\int d^{d}x~\sqrt{-g}  \left(R + O(L^2 R^2)\right) + I_{\text{CFT}}
\end{align}
where $x$ denotes brane coordinates and $O(L^2 R^2)$ schematically denotes higher derivative corrections and $I_{\text{CFT}}$ denotes the (non-local) action of the holographic CFT. Here,
\begin{align}
G_d \sim \frac{G_{d+1}}{L}.
\end{align}
Combined with $L^{d-1} /G_{d+1}\sim c$, where $c$ denotes the CFT's effective number of degrees of freedom, this gives
\begin{align}
cG_d \sim L^{d-2}.
\end{align}
Therefore, $L$ is the scale of the breakdown of the semiclassical expansion on the brane. 

\begin{figure}
\includegraphics[width=.4\textwidth]{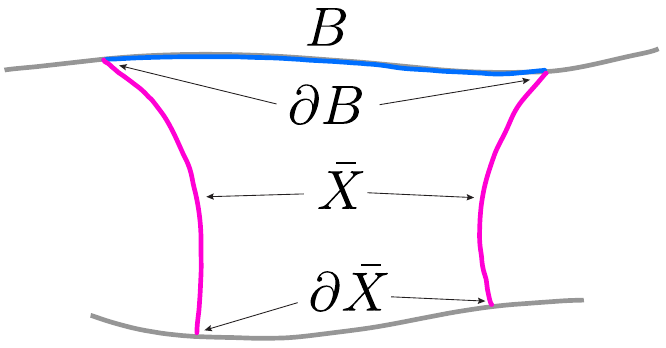}
\caption{An arbitrary region $B$ on the (grey) brane is shown with a minimal area extremal surface $\bar{X}$ homologous to it. While some connected components of $\bar{X}$ end on $\partial B$, in general there may be others which end on another brane (satisfying a Neumann boundary condition). In such cases, we can glue the solution to itself across the lower brane, reducing to a scenario with only Dirichlet boundary condition.}\label{fig-braneN}
\end{figure}

For a general discussion, it will be more convenient to consider Riemann normal coordinates in a neighborhood of the brane:
\begin{align}
    ds^2 = dn^2 + g_{ij}(n,x) dx^i dx^j,
\end{align}
where the brane is located at $n=0$. In these coordinates, the brane equation of motion~\eqref{eq-braneeom123} at $n=0$ give
\begin{align}\label{eq-braneeomn}
\partial_n g_{ij}= -\frac{2}{L} g_{ij}
\end{align}
Now, consider a partial Cauchy slice $B$ of the brane spacetime. We have
\begin{align}\label{eq-Sgendef}
    S_{\text{gen}}(B) = \frac{A (\bar{X}(B))}{4 G_{d+1}},
\end{align}
where $A (\bar{X}(B))$ denotes the area of the minimal area bulk extremal surface $\bar{X}(B)$ homologous to $B$~\cite{Hubeny:2012wa, Emparan:2006ni, Emparan:1999wa, Myers:2013lva, Chen:2020uac, Chen:2020hmv, Bousso:2020kmy}. We may view Eq. \eqref{eq-Sgendef} as a definition of $S_{\text{gen}}(B)$ for our purposes, though it must be possible to derive it from an independent definition of $S_{gen}(B)$.  Note that the homology condition here does not necessarily mean $\partial \bar{X} = \partial B$. In general $\partial B \subset \partial \bar{X}$, where some connected components of $\partial \bar{X}$ may end with Neumann boundary conditions on a brane (See Fig.~\ref{fig-braneN}).

A powerful feature of the brane-world scenario is that bulk Einstein gravity induces on the brane, Einstein gravity plus higher derivative corrections to all orders. This is a very convenient setup to study quantities like the generalized entropy and conditions like the restricted QFC which make sense to all orders in the semiclassical expansion (controlled by $L$). Of course, the bulk theory receives both quantum and stringy correction (discussed in subsection~\ref{subsec-quantumcorrections}), which on the brane are interpreted as $1/c$ and inverse coupling corrections respectively.

Before going on, we will comment on connections to previous works. In \cite{Myers:2013lva}, Eq. \eqref{eq-Sgendef} was expanded in small $L$ where it was shown to reproduce the Bekenstein-Hawking entropy for the region $B$ plus quantum corrections and local extrinsic curvature terms on $\partial B$~\cite{Dong:2013qoa}. Furthermore, our setup is close in spirit to the work of~\cite{Koeller:2016aa} which in the standard AdS/CFT setup, used the HRT formula~\cite{Hubeny:2007xt, Lewkowycz:2013nqa} to prove the quantum null energy condition in holographic CFTs (see also later work~\cite{Fu:2017evt, Akers:2017ttv}). The brane setup here is of course different in that it is a gravitational theory. But, in addition, there are two major technical differences with~\cite{Koeller:2016aa}. First, contrary to~\cite{Koeller:2016aa}, we do not analyze the extremal surface $\bar{X}$ in a ``near boundary/brane'' expansion. The treatment is fully non-perturbative in that regard, enabling us to draw conclusions which hold to all orders in the brane semiclassical expansion parameter $L$. Furthermore, in~\cite{Koeller:2016aa}, a crucial inequality was derived from ``entanglement wedge nesting'', proven earlier only in the context of standard AdS/CFT~\cite{Wall:2012uf}. We therefore use another, more direct, technique here.

\subsection{The Proof}\label{sec:proof}

Let $\bar{X}^\mu(n, y^a)$ specify embedding coordinates for $\bar{X}$  where $\mu$ denotes bulk coordinates and $y^a$ is an extension of the coordinates on $\partial B$ to the bulk. We work in a gauge where $\bar{X}^n=n$. Therefore, $\bar{X}^\mu = (n,\bar{X}^i(n,y^a))$ such that $\bar{X}^i(n=0,y^a)$ is the embedding coordinates of the $\partial B$ on the brane. For simplicity we can pick coordinates on the brane such that $\bar{X}^i(n=0,y^a) =y^a \delta_{ia}$ (no summation). Let $\bar{H}_{\alpha \beta}(n,y^a)$ denote the induced metric on $\bar{X}^\mu$.~\footnote{To sum up the notation, $\mu$ and $\nu$ are bulk indices, $i$ and $j$ are brane indices, $\alpha$ and $\beta$ denote indices along $\bar{X}$, while $a$ denotes indices on $\partial B$. So, e.g. $\mu = \{n, i \}$ and $\alpha = \{n, a \}$.} Here $\alpha$ and $\beta$ are either $n$ or $y^a$, coordinates on $\bar{X}$. Then,
\begin{align}
A[\bar{X}] = \int dn dy^a \sqrt{\bar{H}},
\end{align}
where $\bar{H} = \det(\bar{H}_{\alpha \beta})$. By taking a functional derivative of this area subject to null deformation of the brane region in the $k^i$ or $\ell^i$ directions, we can compute the corresponding quantum expansions,
\begin{align}
    \Theta_{(k)}(B;y) = \left.-\frac{k_\mu t^\mu}{\ell_S}\right\rvert_{\partial B}\label{eq-RSexpansion},\\
    \Theta_{(\ell)}(B;y) = \left.-\frac{\ell_\mu t^\mu}{\ell_S}\right\rvert_{\partial B}\label{eq-RSexpansion2},
\end{align}

where $B$ in the argument of $\Theta$ means evaluating it at $V=0$. Further, $k^\mu$ and $\ell^\mu$ are the push-forwards of $k^i$ and $\ell^i$, $t^\mu$ is the unique unit-normalized tangent vector of $\bar{X}$ orthogonal to $\partial B$, and $\ell_S = G_{d+1}/G_d$ which is defined for convenience. Then, $\ell_S \sim L$, the effective short-distance cutoff of the local gravitational theory on the brane. Therefore, we take appropriate gauge-invariant lengthscales associated to the background spacetime, state, and region $B$ to be much larger than $\ell_S$ to respect the semiclassical regime. Note that since $\bar{X}$ is extremal, the only contribution to Eqs. $\eqref{eq-RSexpansion}$ and $\eqref{eq-RSexpansion2}$ come from the subset of $\partial \bar{X}$ with Dirichlet boundary condition, i.e. $\partial B$.

In our gauge, we have $\bar{H}_{na}\rvert_{\partial B}=0$, a condition which we can preserve in a neighborhood of $\partial B$ on $\bar{X}$ by defining the extension of the $y^a$ coordinates into the bulk appropriately. We also have
\begin{align}
\bar{H}_{nn}(n, y) = 1+ g_{ij} \partial_n\bar{X}^i \partial_n\bar{X}^j.
\end{align}
To make sure that $\bar{X}$ is a spacelike surface, we need $\bar{H}_{nn}>0$. Using Eq. \eqref{eq-RSexpansion}, one can write $\partial_n \bar{X}^i$ in terms of the quantum expansions of $B$,
\begin{align}\label{eq-forapp}
\left. \partial_n \bar{X}^i \right\rvert_{n=0}= -\frac{\ell_S\Theta_{(\ell)}}{\sqrt{1-2 \ell_S^2 \Theta_{(k)}\Theta_{(\ell)}}} k^i - \frac{\ell_S\Theta_{(k)}}{\sqrt{1-2 \ell_S^2 \Theta_{(k)}\Theta_{(\ell)}}} \ell^i,
\end{align}
where $\Theta_{(\ell)}$ is defined in the obvious analogous way for the null vector field orthogonal to $\partial B$ and normalized by $k^i \ell_i = 1$. Then,
\begin{align}
\left.H_{nn}\right\rvert_{n=0}>0 \implies 2\ell_S^{2} \Theta_{(k)}\Theta_{(\ell)}<1.
\end{align}
In fact, we expect (and henceforth assume) from the validity of the semiclassical expansion that
\begin{align}\label{eq-klsmall}
|\Theta_{(k)} \Theta_{(\ell)}| \ll \ell_S^{-2}.
\end{align}
This makes sense because $\Theta_{(k)} \Theta_{(\ell)}$ is a coordinate invariant quantity related to the brane region and state. For example, if we take $B$ to be the ball of radius $R$ in Minkowski space, this condition is equivalent to $R \gg \ell_S$ which is clearly required for the validity of the semiclassical analysis. From now on, we add condition \eqref{eq-klsmall} to the list of other curvature invariants which satisfy the semiclassical condition.

Let $V_{\lambda=0}(y)=0$. Without loss of generality, we focus on conditions \eqref{eq-weakQFC} and \eqref{eq-weakQFCell} when evaluated at $\lambda = 0$. In order to compute $\partial_{\lambda} \Theta_{(k)}(V_{\lambda})\rvert_{\lambda=0}$ and $\partial_{\lambda} \Theta_{(\ell)}(V_{\lambda})\rvert_{\lambda=0}$, we need to calculate the response of the extremal surface $\bar{X}(B)$ to an infinitesimal deformation of $B$ at $\partial B$ in the $k^i$ direction. A deformation of $\bar{X}$ can be specified by a deformation vector field $\alpha \bar{k}^\mu + \beta \bar{\ell}^\mu$ in the normal bundle of $\bar{X}$ where $\bar{k}^\mu$ and $\bar{\ell}^\mu$ are null vector fields orthogonal to $\bar{X}$ (which we normalize with $\bar{k}^\mu \bar{\ell}_\mu =1$), and $\alpha$ and $\beta$ are scalar functions on $\bar{X}$. To deform $\bar{X}$, we can then follow (by a fixed affine parameter $\lambda$) geodesics fired from $\bar{X}$ along $\alpha \bar{k}^\mu + \beta \bar{\ell}^\mu$. After some computation from Eq. \eqref{eq-RSexpansion}, we get
\begin{align}
\partial_{\lambda} \Theta_{(k)}(V_\lambda;y) &= \frac{1}{\ell_S (\bar{H}_{nn})^\frac{1}{2}} \biggl(-k_i \partial_n (\alpha \bar{k}^i + \beta \bar{\ell}^i)\rvert_{n=0}\nonumber\\
&+ \ell_S^3 (\bar{H}_{nn})^{\frac{3}{2}} \Theta_{(k)}\partial_\lambda (\Theta_{(k)} \Theta_{(\ell)}) \biggr). \label{eq-thetakprimegeneral}
\end{align}

\begin{figure}
\includegraphics[width=.4\textwidth]{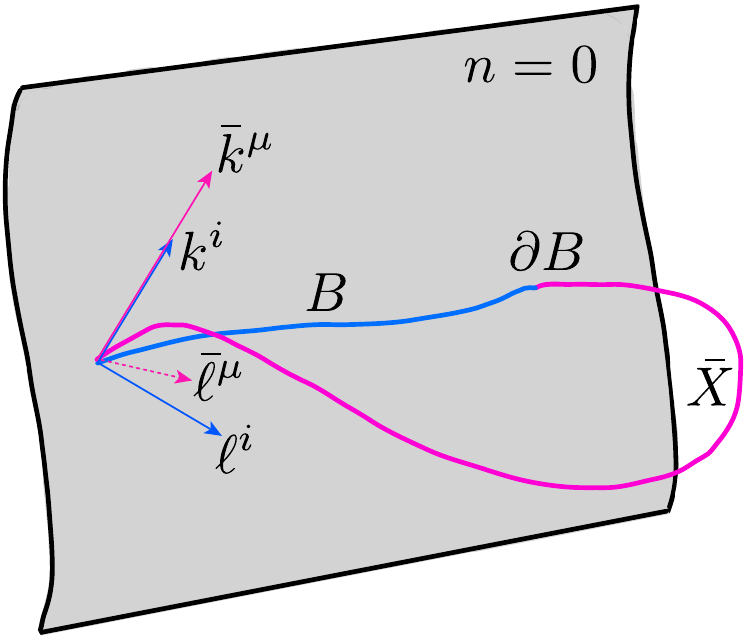}
\caption{$B$ is a subregion on the brane (located at $n=0$). The generalized entropy of $B$ is computed by the Bekenstein-Hawking entropy of the minimal area bulk extremal surface $\bar{X}$ homologous to $B$. The null vector fields $k^i$ and $\ell^i$ orthogonal to $\partial B$ and the null vector fields $\bar{k}^\mu$ and $\bar{\ell}^\mu$ orthogonal to $\partial \bar{X}$ are depicted. $k^i$ and $\bar{k}^\mu$ align at a point $y\in \partial B$ where $\Theta_{(k)}(B;y)=0$.}\label{fig-brane}
\end{figure}

At $\partial B$, the deformation of $\bar{X}$ projected onto the brane needs to satisfy the following condition:
\begin{align}
(\alpha \bar{k}^i + \beta \bar{\ell}^i)\rvert_{\partial B} = (\partial_\lambda V_\lambda) k^i,
\end{align}
where $\bar{k}^i$ and $\bar{\ell}^i$ are the projections of $\bar{k}^\mu$ and $\bar{\ell}^\mu$ onto the brane. In general, $\bar{k}^i$ ($\bar{\ell}^i$) is different from $k^i$ ($\ell^i$). See Fig. \ref{fig-brane}.

Using the definitions of $\bar{k}^\mu$ and $\bar{\ell}^\mu$, it is possible to derive
\begin{align}
\bar{k}^i\rvert_{\partial B} &= k^i + \frac{-1+ \ell_S^2 \Theta_{(k)}\Theta_{(\ell)} +\sqrt{1-2 \ell_S^2 \Theta_{(k)}\Theta_{(\ell)}} }{\ell_S^2 \Theta_{(\ell)}^2} \ell^i, \\
\bar{\ell}^i\rvert_{\partial B} &=\frac{1-\ell_S^2 \Theta_{(k)}\Theta_{(\ell)}+\sqrt{1-2\ell_S^2 \Theta_{(k)}\Theta_{(\ell)}}}{2} \ell^i \nonumber\\
&-\frac{\ell_S^2 \Theta_{(\ell)}^2}{2} k^i.
\end{align}
From this, we can derive
\begin{align}
\alpha\rvert_{\partial B}&=\frac{1-\ell_S^2 \Theta_{(k)}\Theta_{(\ell)}+\sqrt{1-2\ell_S^2 \Theta_{(k)}\Theta_{(\ell)}}}{2\sqrt{1-2 \ell_S^2 \Theta_{(k)}\Theta_{(\ell)}}} \partial_\lambda V_{\lambda},\label{eq-alpha1}\\
\beta\rvert_{\partial B} &=\frac{1-\ell_S^2 \Theta_{(k)}\Theta_{(\ell)}-\sqrt{1-2\ell_S^2 \Theta_{(k)}\Theta_{(\ell)}}}{\ell_S^2 \Theta_{(\ell)}^2\sqrt{1-2 \ell_S^2 \Theta_{(k)}\Theta_{(\ell)}}} \partial_\lambda V_{\lambda} \label{eq-beta1}.
\end{align}
We can simplify the above expressions using only $\ell_S^2 \Theta_{(k)}\Theta_{(\ell)} \ll 1$:
\begin{align}
\bar{k}^i\rvert_{\partial B} &= k^i - \frac{\ell_S^2 \Theta_{(k)}^2}{2} \ell^i +\cdots,\\
\bar{\ell}^i\rvert_{\partial B} &= \ell^i -\frac{\ell_S^2 \Theta_{(\ell)}^2}{2} k^i + \cdots,
\end{align}
And
\begin{align}
\alpha\rvert_{\partial B}&=\partial_\lambda V_{\lambda}+\cdots, \label{eq-alphapert}\\ 
\beta\rvert_{\partial B} &=\frac{\ell_S^2 \Theta_{(k)}^2}{2}\partial_\lambda V_{\lambda} + \cdots. \label{eq-betapert}
\end{align}
As an important side note, it does not make sense to demand that the absolute values of $\ell_S \Theta_{(k)}$ and $\ell_S \Theta_{(\ell)}$ are small because their values can change under a simultaneous rescaling of $k^i$ and $\ell^i$. In other words, these dimensionless quantities are coordinate dependent.\footnote{For example, for an evaporating black hole in infalling Eddington-Finkelstein coordinates, it is possible to make them arbitrarily large~\cite{Penington:2019npb, Almheiri:2019psf}.}

As functions on $\bar{X}$, $\alpha$ and $\beta$ are constrained by the fact that the deformation of $\bar{X}$ needs to take it to a nearby extremal surface. Deriving this constraint is a straightforward exercise (see e.g.~\cite{Engelhardt:2019hmr}). The result is
\begin{align}\label{eq-elliptic}
    \begin{pmatrix}
         \hat{D}_+ & -\bar{\varsigma}_{(\bar{\ell})}^2-8 \pi G \bar{T}_{\mu\nu} \bar{\ell}^\mu \bar{\ell}^\nu \\ 
         -\bar{\varsigma}_{(\bar{k})}^2-8 \pi G \bar{T}_{\mu\nu} \bar{k}^\mu \bar{k}^\nu & \hat{D}_-
     \end{pmatrix}
     \begin{pmatrix}
          \alpha\\ 
          \beta  
     \end{pmatrix}
      =
      0,
\end{align}
where $\bar{\varsigma}_{(k)}^2$ and $\bar{\varsigma}_{(\ell)}^2$ denote shear-squared terms on $\bar{X}$, and
\begin{align}
\hat{D}_{\pm} = -\bar{\nabla}^2 \mp 2 \chi^\alpha \bar{\nabla}_\alpha -\left(\bar{\chi}^\alpha \bar{\chi}_\alpha \pm \bar{\nabla}_\alpha \bar{\chi}^\alpha + \bar{G}_{\mu\nu} \bar{k}^\mu \bar{\ell}^\nu-\frac{\bar{r}}{2}\right),
\end{align}
where $\bar{\nabla}_\alpha$ is the covariant derivative on $\bar{X}^\mu$, $\bar{\chi}_\alpha = \bar{\ell}^\mu \bar{\nabla}_\alpha \bar{k}_\mu$, $\bar{G}_{\mu\nu}$ is the bulk Einstein tensor, and $\bar{r}$ is the intrinsic Ricci scalar on $\bar{X}$.

The matrix in Eq. \eqref{eq-elliptic} is a particular linear operator acting on pairs of scalar functions on $\bar{X}$. The result is a ``cooperative elliptic system'' which has in particular been studied in~\cite{Sweerselliptic} and was first discussed in the context of standard AdS/CFT correspondence in~\cite{Engelhardt:2019hmr}. We will use an important theorem in these works, a special case of which (adapted to our needs) we state here:
\begin{thm}\label{thm1}
Consider a fully coupled cooperative elliptic system, i.e., a system of linear differential equations
\begin{align}\label{eq-ellipticthm}
    \begin{pmatrix}
         \hat{L}_1 & f \\ 
         g & \hat{L}_2
     \end{pmatrix}
     \begin{pmatrix}
          A\\ 
          B  
     \end{pmatrix}
      =
      0,
\end{align}
where $A$ and $B$ are functions on an open domain $U$ of $\mathbb{R}^n$, $f$ and $g$ are non-positive functions, and $L_i$ (for $i=1$ or $2$) are elliptic operators
\begin{align}
\hat{L}_i = (H_{i})^{\alpha \beta} \partial_\alpha \partial_\beta + (b_i)^\alpha \partial_\alpha+ c_i,
\end{align}
with $(H_{i})^{\alpha \beta}$ positive-definite matrices for each $i$.

Now, suppose Eq. \eqref{eq-ellipticthm} has a supersolution $(A^+, B^+)$, i.e.,
\begin{align}
A^+\rvert_U \geq 0\\
B^+\rvert_U \geq 0\\
(\hat{L}_1 A^+ + f B^+)\rvert_U \geq 0\label{eq-96}\\
(\hat{L}_2 B^+ + g A^+)\rvert_U \geq 0\label{eq-69}
\end{align}
and either $A^+$ or $B^+$ is non-zero somewhere on $\partial U$, or either \eqref{eq-96} or \eqref{eq-69} is not saturated somewhere in $U$.

Then, for \emph{any} (sufficiently smooth) solution $(A, B)$ to Eq. \eqref{eq-ellipticthm}, either
\begin{align}\label{eq-thmineq}
    \begin{cases}
        A \rvert_{\partial U} \geq 0\\
        B \rvert_{\partial U} \geq 0
    \end{cases} 
    \implies
    \begin{cases}
        A \rvert_{U} > 0\\
        B \rvert_{U} > 0
    \end{cases}
\end{align}
or
\begin{align}
     \begin{cases}
        A \rvert_{U} = 0\nonumber\\
        B \rvert_{U} = 0
     \end{cases}
\end{align}

\end{thm}

In~\cite{Engelhardt:2019hmr}, Theorem \ref{thm1} was applied to the extremal surface deviation Eq. \eqref{eq-elliptic} in the standard AdS/CFT context. We assume that the extension of this theorem from domains of $\mathbb{R}^n$ to a manifold like $\bar{X}$ is trivial. The bulk null energy condition implies\footnote{Alternatively, we can simply assume the (classical) restricted focusing in the bulk.}
\begin{align}\label{eq-ellipticassump1}
\left.(-\bar{\varsigma}_{(\bar{k})}^2-8 \pi G \bar{T}_{\mu\nu} \bar{k}^\mu \bar{k}^\nu)\right\rvert_{\bar{X}} \leq 0 ,\\
\left.(-\bar{\varsigma}_{(\bar{\ell})}^2-8 \pi G \bar{T}_{\mu\nu} \bar{\ell}^\mu \bar{\ell}^\nu)\right\rvert_{\bar{X}}\leq 0.
\end{align}
In the highly non-generic case where one of the above inequalities in saturated everywhere on $\bar{X}$, the analysis becomes trivial. Therefore, without losing anything, we restrict to the case where they are both non-saturated somewhere on $\bar{X}$. Then, the only remaining step to make Theorem~\ref{thm1} non-trivially applicable is to demonstrate the existence of a supersolution. In the standard AdS/CFT context, this follows from the (classical) maximin prescription~\cite{Wall:2012uf}.

This brings us to our main technical assumption: in our setup, where the bulk is cut off by a brane, we will henceforth \emph{assume} that such a supersolution exists. We leave a proof of this assumption to future work, but we comment here on why we believe this is a mild assumption. It is possible to prove that the matrix operator in Eq. \eqref{eq-elliptic} has a real eigenvalue (called the principal eigenvalue) which is equal to or smaller than the real part of all of its other eigenvalues, and whose corresponding eigenvector is a pair of positive functions on $\bar{X}$~\cite{Arvin:2023}. The central assumption here would then follow if this eigenvalue is positive. In the standard AdS/CFT setup, the positivity of this eigenvalue is a simple consequence of the (classical) maximin prescription. Now, from Eq. \eqref{eq-smallz}, we expect that for an $\bar{X}$ anchored to the brane, this eigenvalue is only perturbatively (in $L/L_0$) different from that of standard AdS/CFT, therefore maintaining its positive sign.

Lastly, if there exists connected components of $\partial \bar{X}$ satisfying Neumann boundary conditions on some brane, we can ``double-up'' the solution by gluing across the brane, which would then reduce the boundary conditions of $\bar{X}$ to purely Dirichlet ones (See Fig.~\ref{fig-braneN}).

Assuming the existence of a supersolution, it follows that:
\begin{align}\label{eq-alphabeta}
    \begin{cases}
        \alpha \rvert_{\partial B} \geq 0\\
        \beta \rvert_{\partial B} \geq 0
    \end{cases} 
    \implies
    \begin{cases}
        \alpha \rvert_{\bar{X}} \geq 0\\
        \beta \rvert_{\bar{X}} \geq 0
    \end{cases}
\end{align}

Armed with \eqref{eq-alphabeta}, we can now prove our main results, conditions \eqref{eq-weakQFC} and \eqref{eq-weakQFCell}. First, since $\partial_\lambda V_\lambda \geq 0$, Eqs. \eqref{eq-alphapert} and \eqref{eq-betapert} imply the LHS of \eqref{eq-alphabeta}. Furthermore, $\Theta_{(k)}=0$ simplifies Eq. \eqref{eq-thetakprimegeneral} in the following way:

\begin{align}
	&\Theta_{(k)}(B;y)=0\nonumber\\
	 &\implies \partial_{\lambda} \Theta_{(k)}(V_\lambda;y)\rvert_{\lambda=0} = -\left.\frac{\alpha k_i \partial_n \bar{k}^i + \partial_n \beta}{\ell_S}\right\rvert_{n=0}.
\end{align}
By Eq. \eqref{eq-beta1}, $\beta(n=0,y)=0$. Then, Eq. \eqref{eq-alphabeta} implies that:
\begin{align}\label{eq-signdnbeta}
    \partial_n \beta(n,y)\rvert_{n=0} \geq 0.
\end{align}
To make contact with $k_i\bar{\ell}^i$, we first use $\bar{k}^\mu \bar{k}_\mu=0$:
\begin{align}
    (\bar{k}^n)^2 + g_{ij} \bar{k}^i \bar{k}^j=0
\end{align}
where $\bar{k}^n$ is the component of $\bar{k}^\mu$ orthogonal to the brane. Note that $\Theta_{(k)}(B;y)=0$ implies $\bar{k}^i(n=0,y)=k^i$ which in turn implies $\bar{k}^n(n=0,y)=0$. Taking an $n$ derivative results in:
\begin{align}\label{eq-appref}
 g_{ij} k^i \partial_n\bar{k}^j(n,y)\rvert_{n=0}=0
\end{align}
where we made use of the brane equations of motion $\partial_n g_{ij}\rvert_{n=0} \propto g_{ij}$, $\bar{k}^i(n=0,y)=k^i$, and $\bar{k}_n \partial_{n}\bar{k}^n(n,y)\rvert_{n=0}=0$. The last condition requires showing that $\partial_n \bar{k}^n$ is convergent enough as $n\to 0$ which we have relegated to Appendix~\ref{app-kl}. Putting everything together,
\begin{align}\label{eq-restrictedQFCfinal}
	&\Theta_{(k)}(B;y)=0\nonumber\\
	 &\implies \partial_{\lambda} \Theta_{(k)}(V_\lambda;y)\rvert_{\lambda=0} = -\frac{\partial_n \beta(n,y)\rvert_{n=0}}{\ell_S}\leq0
\end{align}
This concludes the proof of \eqref{eq-weakQFC}. Let us emphasize the role played by the condition $\Theta_{(k)}=0$ in the restricted QFC. It implies $\beta(n=0,y)=0$, which by Eq. \eqref{eq-alphabeta} leads to $\partial_n \beta(n,y)\rvert_{n=0}\geq 0$, something crucial in deriving the bound in Eq. \eqref{eq-restrictedQFCfinal}. Without it, we were unable to prove a bound on $\partial_\lambda \Theta_{(k)}(V_\lambda)$. However, in subsection~\ref{sec:appqfc}, we discuss a concrete sense in which $\beta(n,y) \geq 0$ ``approximately'' bounds how positive $\partial_\lambda \Theta_{(k)}(V_\lambda;y)$ can get when $\Theta_{(k)}(V_\lambda;y)\neq 0$.

For condition \eqref{eq-weakQFCell}, we have:
\begin{align}\label{eq-4531}
	&\Theta_{(\ell)}(B;y)=0\nonumber\\
	 &\stackrel{\partial_\lambda V_\lambda(y)=0}{\implies} \partial_{\lambda} \Theta_{(\ell)}(V_\lambda;y)\rvert_{\lambda=0} = -\frac{\partial_n \alpha(n,y)\rvert_{n=0}}{\ell_S}\leq0.
\end{align}
The inequality in \eqref{eq-4531} follows from
\begin{align}
\partial_\lambda V_\lambda(y)=0 \implies \alpha(n=0,y)=0 \implies \partial_n \alpha(n,y)\rvert_{n=0} \geq 0,
\end{align}
where the second implication follows from condition \eqref{eq-alphabeta}.

In the remainder of this section, we discuss two additional inequalities which follow from strong-subadditivity of the von Neumann entropy (See footnote 3 and~\cite{Wall:2009wi, Bousso:2015eda}). These are

\begin{align}
\left.\frac{\delta \Theta_{(k)}(V;y)}{\delta V(y')}\right\rvert_{y\neq y'} \leq 0,\label{eq-SSA11}\\
\left.\frac{\delta \Theta_{(\ell)}(V;y)}{\delta V(y')}\right\rvert_{y\neq y'} \leq 0\label{eq-SSA22}.
\end{align}
Deriving these conditions is a nice consistency check. This can be done by choosing $V_{\lambda} = \lambda \delta^{d-2} (y-y')$. Then, for $y \neq y'$:
\begin{align}
\left.\frac{\delta \Theta_{(k)}(V;y)}{\delta V(y')}\right\rvert_{V=0} &= \frac{-1}{\ell_S}\biggl(\partial_n\beta(n,y)\rvert_{n=0}\nonumber\\
&+\frac{\ell_S^2 \Theta_{(k)}(B;y)^2}{2} \partial_n \alpha(n,y)\rvert_{n=0}\biggr),\label{eq-112} \\
\left.\frac{\delta \Theta_{(\ell)}(V;y)}{\delta V(y')}\right\rvert_{V=0} &= \frac{-1}{\ell_S}\biggl(\partial_n \alpha(n,y)\rvert_{n=0}\nonumber\\
&+ \frac{\ell_S^2 \Theta_{(\ell)}(B;y)^2}{2} \partial_n\beta(n,y)\rvert_{n=0}\biggr).\label{eq-113}
\end{align}
where we have dropped terms suppressed by $\ell_S^2\Theta_{k} \Theta_{\ell}$, which are not relevant since they multiply either $\partial_n\beta$ or $\partial_n\alpha$ in the expressions above. Now, condition \eqref{eq-alphabeta} implies that for $y\neq y'$,
\begin{align}
\partial_n \alpha (n,y)\rvert_{n=0} &\geq 0,\\
\partial_n \beta (n,y)\rvert_{n=0} &\geq 0,
\end{align}
resulting in the desired signs \eqref{eq-SSA11} and \eqref{eq-SSA22}.

\section{Discussion}\label{sec:disc}

The following ideas will be explored and expanded on in forthcoming work.

\subsection{Bulk quantum and higher curvature corrections}\label{subsec-quantumcorrections}

Even though Eq. \eqref{eq-Sgendef} already includes all perturbative in $G_{d}$ corrections, it receives additional \emph{bulk} quantum, i.e., $O(G_{d+1})$, and higher curvature corrections, i.e., $O(\delta)$, where $\delta$ is the small scale suppressing the higher curvature terms in the bulk gravity action. Studying these corrections (which are $1/c$ and inverse coupling corrections from the brane perspective) is very important since it will elucidate whether restricted QFC (or at least its proof here) is an accident of a leading order analysis of the brane-world or something which holds more generally. We comment on how one could extend the proof of restricted QFC to include these corrections, leaving a thorough analysis to future work. Following the quantum extremal surface prescription~\cite{Engelhardt:2014aa}, we assume that the exact formula, i.e., to all orders in bulk perturbation theory, for the brane generalized entropy is given by:
\begin{align}
    S_{\text{gen}}(B) = S_{\text{gen}}(\bar{H}(B))
\end{align}
where $\bar{H}(B)$ is the homology slice of the quantum extremal surface $\bar{X}(B)$ homologous to $B$ with the smallest bulk generalized entropy. Here the bulk generalized entropy is given by:
\begin{align}
    S^{\text{bulk}}_{\text{gen}}(\bar{H}(B)) = Q(\bar{X}(B)) + S^{\text{bulk}}(\bar{H}(B)) + \cdots
\end{align}
where $Q(\bar{X})$ is the Dong entropy functional~\cite{Dong:2013qoa}:
\begin{align}
    Q(\bar{X}) = \frac{A(\bar{X})}{4 G_{d+1}} + O(\lambda) 
\end{align}
and $S^{\text{bulk}}(\bar{H}(B))$ denotes the bulk von Neumann entropy in $H(B)$.

For proving \eqref{eq-weakQFC} and \eqref{eq-weakQFCell}, these perturbative corrections only matter if Eq. \eqref{eq-signdnbeta} is saturated at leading order. For condition \eqref{eq-weakQFC} saturation implies:
\begin{align}
\partial_n \beta(n,y)\rvert_{n=0}=0,
\end{align}
a generalization of Hopf Lemma~\cite{renardy2006introduction} then implies the very stringent condition that
\begin{align}\label{eq-}
\beta\rvert_{\bar{X}}=0,
\end{align}
That is, at leading order a small null deformation of $\partial B$ in the null direction generates a null deformation \emph{everywhere} on $\bar{X}$. Inspecting the extremal surface deviation Eq. \eqref{eq-elliptic}, this also implies that to leading order (in $\delta$ or $G_{d+1}$), $\bar{X}$ lies on a locally stationary horizon. This simplifies the analysis greatly. The next-to-leading order corrections can be solved for explicitly by Eq. \eqref{eq-elliptic}. A possibility is that $\partial_\lambda \Theta_{(k)}(V_\lambda;y)$ reduces at next-to-leading order to integrated bulk restricted QFC. Higher order corrections will not be important if the saturation of the integrated bulk restricted QFC is only possible it is so to all orders in the bulk $\delta$ or $G_{d+1}$ expansions. A similar argument can be made for condition \eqref{eq-weakQFCell}.

One could also consider a generalization of our setup where additional intrinsic brane curvature terms (beyond the pure tension term) are added directly to the brane action~\cite{Dvali:2000hr}. If such terms are perturbatively small, i.e., they only cause small changes to the coefficients of the brane gravity derivative expansion, Eq. \eqref{eq-branederivexp}, then it is possible that the treatment discussed earlier in this subsection would suffice to generalize the proofs of restricted QFC. If such corrections are large though, we do not how to use our method to derive the restricted QFC. One possibility is that such theories are pathological. This possibility was discussed in~\cite{Leichenauer:2018tnq}, where the authors emphasized that non-tension terms added to the brane lead to a brane null geodesic not being a bulk null geodesic (since $K_{ij}$ will no longer be proportional to $g_{ij}$), therefore violating the ``brane causality condition'', i.e., there will be bulk causal curves connecting points on the brane which are spacelike separated on the brane's causal structure.

\subsection{Approximate QFC}\label{sec:appqfc}

By Taylor expanding $\beta$ near $n=0$, we get (modulo O(1) factors in the coefficients)
\begin{align}
\beta(n,y) &\sim \ell_S^2 \Theta^2_{(k)} - n \ell_S(\partial_\lambda \Theta_{(k)} + \Theta^2_{(k)})\nonumber\\
&+ n^2 \partial_n^2 \beta(n,y)\rvert_{n=0}+ O(n^3).
\end{align}
From Eq. \eqref{eq-elliptic}, we have that $\partial_n^2 \beta(n,y)\rvert_{n=0} \sim \ell_S^{-2}$. Therefore, if at some value of $n$, the first two terms become equal while their absolute values are much larger than the third and higher order terms, then $\beta(n,y) \geq 0$ would be violated. It is easy to check that this leads to an ``approximate quantum focusing'' condition\footnote{Douglas Stanford suggested a similar bound during a discussion about this work.}
\begin{align}\label{eq-approxQFC}
\partial_\lambda \Theta_{(k)} \lesssim  (\partial_\lambda V_\lambda) \Theta_{(k)}^2.
\end{align}
This bound becomes a sharp statement when there exists a perturbative parameter $\epsilon$ in the problem and $\partial_\lambda \Theta_{(k)}$ and $\Theta_{(k)}$ acquire $\epsilon$ expansions. Then, in the $\epsilon \to 0$ limit, \eqref{eq-approxQFC} states that the leading LHS term, if it is of lower order in $\epsilon$ than the leading RHS term, is non-positive.

It would be interesting to explore non-trivial applications of \eqref{eq-approxQFC}. Here we provide one. In~\cite{Wall:2015raa}, it was found that in Einstein gravity plus higher curvature corrections, classical focusing (of the Dong entropy functional~\cite{Dong:2013qoa}) is upheld on cross-sections of a causal horizon which is a slight perturbation of a Killing horizon. This was shown by observing that $\partial_\lambda \Theta_{(k)} = -G_d (\partial_\lambda V_\lambda) T_{ij} k^i k^j + O(G_d^2)$, which is then non-positive at $O(G_d)$ by the null energy condition.

While this does not follow from the restricted QFC (since $\Theta_{(k)}$ is generally non-zero on such perturbed horizons), it does follow from \eqref{eq-approxQFC}: on the perturbed horizon, $\Theta_{(k)} = O(G_d)$, forcing any leading term in $\partial_\lambda \Theta_{(k)}$ lower than $O(G_d^2)$ to be non-positive.

\subsection{Does a QFC counter-example exist?}

As discussed earlier, while restricted quantum focusing \eqref{eq-weakQFC} has a natural proof in the brane-world scenario, it is not clear to us how to leverage the same technique to prove the original QFC \eqref{eq-QFC}. This begs the question of whether the QFC is true.

Here we discuss a setup where a QFC counter-example may be plausible. By Raychaudhuri's equation in Einstein gravity, we have
\begin{align}\label{eq-1234}
    \Theta'_{(k)} =
    -\frac{\theta^2_{(k)}}{d-2} - \varsigma^2_{(k)} -4 G_{d} (2\pi \langle \hat{T}_{ij}\rangle k^i k^j-\hat{S}_{(k)}''),
\end{align}
where $\theta_{(k)}$ and $\varsigma^2_{(k)}$ are the classical expansion and the shear-squared of $\partial B$ respectively, $\langle \hat{T}_{ij}\rangle$ is the expectation value of the renormalized stress-energy tensor, and
\begin{align}
\Theta'_{(k)} \delta^{d-2} (y-y') = \lim_{V \to \lambda \delta^{d-2} (y-y')} \partial_\lambda \Theta_{(k)}(V_\lambda;y),\\
\hat{S}_{(k)}'' \delta^{d-2} (y-y')= \lim_{V \to \lambda \delta^{d-2} (y-y')} \partial_\lambda \left(\left.\frac{1}{\sqrt{h_V}}\frac{\delta \hat{S}}{\delta V}\right\rvert_{V_\lambda}\right),
\end{align}
where $\hat{S}$ denotes the renormalized von Neumann entropy of bulk fields.
In~\cite{Leichenauer:2018obf, Balakrishnan:2019gxl}, substantial evidence was provided that for interacting CFTs, at least when the domain of dependence of $B$ is a Rindler wedge, we have
\begin{align}\label{eq-qnecsat1}
2\pi \langle \hat{T}_{ij} \rangle k^i k^j = \hat{S}''_{(k)}.
\end{align}

Now, let $B$ be a ball in flat space. We then expect new terms in Eq. \eqref{eq-qnecsat1}. In particular, by dimensional analysis we expect a term proportional to $\theta_{(k)} \hat{S}_{(k)}'$. Such a term does not have a definite sign and, when $\theta_{(k)}/ S'_{(k)} = O(G_d)$, its sign may affect the sign of $\Theta'_{(k)}$.

Interestingly, when we instead consider the restricted QFC, where we have the additional constraint $\Theta_{(k)}=0$, i.e.,
\begin{align}
\Theta_{(k)}= \theta_{(k)}+ 4 G_d \hat{S}_{(k)}' = 0,
\end{align}
then $\theta_{(k)} \hat{S}_{(k)}'$ does acquire a definite sign, giving the restricted QFC a fighting chance. Examples like this will be explored in forthcoming work.

\section*{Acknowledgements}
I would like to thank Chris Akers, Raphael Bousso, Ven Chandrasekaran, Netta Englehardt, Tom Hartman, Adam Levine, Raghu Mahajan, Pratik Rath, Douglas Stanford, Leonard Susskind, and Zhenbin Yang for discussions. I am especially grateful to Netta Engelhardt and Adam Levine for providing useful feedback on an early draft of this paper. This work was supported by the National Science Foundation under Award Number 2014215.

\appendix

\section{Justifying Eq. \eqref{eq-appref}}\label{app-kl}
The bulk metric in a neighborhood of the brane is given by
\begin{align}
    ds^2 = dn^2 + g_{ij}(n,x) dx^i dx^j.
\end{align}

The vector fields $\bar{k}^\mu$ is orthogonal to $\bar{X}$ and null, so in particular 
\begin{align}
    \bar{k}^n + g_{ij} \bar{k}^i \partial_n \bar{X}^j=0,\\
    (\bar{k}^n)^2 + g_{ij} \bar{k}^{i}\bar{k}^{j}=0.
\end{align}

Taking $n$ derivatives of the above equations, we get
\begin{align}
    &\lim_{n\to0}(\partial_n \bar{k}^n + g_{ij} \partial_n \bar{k}^i \partial_n \bar{X}^j + g_{ij} \bar{k}^i \partial_n^2 \bar{X}^j)=0,\\
    & \lim_{n\to0}(\bar{k}^n \partial_n \bar{k}^n + g_{ij} \partial_n \bar{k}^i \bar{k}^j) =0,
\end{align}
where we used $\Theta_{(k)}(B; y)=0$ and the brane equations of motion to simplify the first expression. By Eq. \eqref{eq-forapp}, a combination of the above equations give
\begin{align}\label{eq-randoapp}
\lim_{n\to0}\left[\partial_n \bar{k}^n \left(1+\frac{\ell_S \Theta_{\ell}~\bar{k}^n}{\sqrt{1-2\ell_S^2 \Theta_{(k)}\Theta_{(\ell)}}}\right)+g_{ij} \bar{k}^i \partial_n^2 \bar{X}^j\right]=0
\end{align}

Since $\Theta_{(k)}(B;y)$ and $\Theta_{(\ell)}(B;y)$ are finite, this implies via Eq. \eqref{eq-forapp} that $\partial_n \bar{X}^i\rvert_{n=0}$ is finite. We now take advantage of the extremal surface equation:
\begin{align}
    \frac{1}{\sqrt{\bar{H}}} \partial_\alpha (\sqrt{\bar{H}} \bar{H}^{\alpha \beta} \partial_\beta \bar{X}^i) + \bar{H}^{\alpha \beta} \bar{\Gamma}^{i}_{k l} \partial_\alpha \bar{X}^k \partial_\beta \bar{X}^l=0
\end{align}

This equation relates $g_{ij} \bar{k}^i \partial_n^2 \bar{X}^j\rvert_{n=0}$ to terms involving $\partial_n\bar{X}^i$. By the smoothness of $\partial B$, we expect $\Theta_{(k)}(B;y)$ and $\Theta_{(\ell)}(B;y)$ to be well-behaved (e.g., finite and differentiable at $y$), which then enforces $g_{ij} \bar{k}^i \partial_n^2 \bar{X}^j\rvert_{n=0}$ to be finite. Since $\bar{k}^n (n=0,y)=0$, Eq. \eqref{eq-randoapp} now implies the desired result that $\lim_{n\to 0}\bar{k}^n(n,y)\partial_n \bar{k}^n(n,y) = 0$.

\bibliographystyle{JHEP}
\bibliography{main, all, all1}

\providecommand{\href}[2]{#2}\begingroup\raggedright\begin{thebibliography}{10}

\bibitem{Bekenstein:1973ur}
J.~D. Bekenstein, \emph{{Black Holes and Entropy}},
  \href{https://doi.org/10.1103/PhysRevD.7.2333}{\emph{Phys. Rev.} {\bfseries
  D7} (1973) 2333}.

\bibitem{tHooft:1984kcu}
G.~'t~Hooft, \emph{{On the Quantum Structure of a Black Hole}},
  \href{https://doi.org/10.1016/0550-3213(85)90418-3}{\emph{Nucl. Phys. B}
  {\bfseries 256} (1985) 727}.

\bibitem{Susskind:1994sm}
L.~Susskind and J.~Uglum, \emph{{Black hole entropy in canonical quantum
  gravity and superstring theory}},
  \href{https://doi.org/10.1103/PhysRevD.50.2700}{\emph{Phys. Rev.} {\bfseries
  D50} (1994) 2700} [\href{https://arxiv.org/abs/hep-th/9401070}{{\ttfamily
  hep-th/9401070}}].

\bibitem{Dong:2013qoa}
X.~Dong, \emph{{Holographic Entanglement Entropy for General Higher Derivative
  Gravity}}, \href{https://doi.org/10.1007/JHEP01(2014)044}{\emph{JHEP}
  {\bfseries 01} (2014) 044} [\href{https://arxiv.org/abs/1310.5713}{{\ttfamily
  1310.5713}}].

\bibitem{Maldacena:1997re}
J.~M. Maldacena, \emph{{The Large N limit of superconformal field theories and
  supergravity}}, \href{https://doi.org/10.1023/A:1026654312961}{\emph{Int. J.
  Theor. Phys.} {\bfseries 38} (1999) 1113}
  [\href{https://arxiv.org/abs/hep-th/9711200}{{\ttfamily hep-th/9711200}}].

\bibitem{Engelhardt:2014gca}
N.~Engelhardt and A.~C. Wall, \emph{{Quantum Extremal Surfaces: Holographic
  Entanglement Entropy beyond the Classical Regime}},
  \href{https://doi.org/10.1007/JHEP01(2015)073}{\emph{JHEP} {\bfseries 01}
  (2015) 073} [\href{https://arxiv.org/abs/1408.3203}{{\ttfamily 1408.3203}}].

\bibitem{Dong:2017aa}
X.~Dong and A.~Lewkowycz, \emph{Entropy, extremality, euclidean variations, and
  the equations of motion},  \href{https://arxiv.org/abs/1705.08453}{{\ttfamily
  1705.08453}}.

\bibitem{Akers:2020pmf}
C.~Akers and G.~Penington, \emph{{Leading order corrections to the quantum
  extremal surface prescription}},
  \href{https://doi.org/10.1007/JHEP04(2021)062}{\emph{JHEP} {\bfseries 04}
  (2021) 062} [\href{https://arxiv.org/abs/2008.03319}{{\ttfamily
  2008.03319}}].

\bibitem{Page:1993wv}
D.~N. Page, \emph{{Information in black hole radiation}},
  \href{https://doi.org/10.1103/PhysRevLett.71.3743}{\emph{Phys. Rev. Lett.}
  {\bfseries 71} (1993) 3743}
  [\href{https://arxiv.org/abs/hep-th/9306083}{{\ttfamily hep-th/9306083}}].

\bibitem{Penington:2019npb}
G.~Penington, \emph{{Entanglement Wedge Reconstruction and the Information
  Paradox}}, \href{https://doi.org/10.1007/JHEP09(2020)002}{\emph{JHEP}
  {\bfseries 09} (2020) 002}
  [\href{https://arxiv.org/abs/1905.08255}{{\ttfamily 1905.08255}}].

\bibitem{Almheiri:2019psf}
A.~Almheiri, N.~Engelhardt, D.~Marolf and H.~Maxfield, \emph{{The entropy of
  bulk quantum fields and the entanglement wedge of an evaporating black
  hole}}, \href{https://doi.org/10.1007/JHEP12(2019)063}{\emph{JHEP} {\bfseries
  12} (2019) 063} [\href{https://arxiv.org/abs/1905.08762}{{\ttfamily
  1905.08762}}].

\bibitem{Hartman:2020swn}
T.~Hartman, E.~Shaghoulian and A.~Strominger, \emph{{Islands in Asymptotically
  Flat 2D Gravity}}, \href{https://doi.org/10.1007/JHEP07(2020)022}{\emph{JHEP}
  {\bfseries 07} (2020) 022}
  [\href{https://arxiv.org/abs/2004.13857}{{\ttfamily 2004.13857}}].

\bibitem{Gautason:2020tmk}
F.~F. Gautason, L.~Schneiderbauer, W.~Sybesma and L.~Thorlacius, \emph{{Page
  Curve for an Evaporating Black Hole}},
  \href{https://doi.org/10.1007/JHEP05(2020)091}{\emph{JHEP} {\bfseries 05}
  (2020) 091} [\href{https://arxiv.org/abs/2004.00598}{{\ttfamily
  2004.00598}}].

\bibitem{Brown:2019rox}
A.~R. Brown, H.~Gharibyan, G.~Penington and L.~Susskind, \emph{{The
  Python\textquoteright{}s Lunch: geometric obstructions to decoding Hawking
  radiation}}, \href{https://doi.org/10.1007/JHEP08(2020)121}{\emph{JHEP}
  {\bfseries 08} (2020) 121}
  [\href{https://arxiv.org/abs/1912.00228}{{\ttfamily 1912.00228}}].

\bibitem{Engelhardt:2021qjs}
N.~Engelhardt, G.~Penington and A.~Shahbazi-Moghaddam, \emph{{Finding pythons
  in unexpected places}},
  \href{https://doi.org/10.1088/1361-6382/ac3e75}{\emph{Class. Quant. Grav.}
  {\bfseries 39} (2022) 094002}
  [\href{https://arxiv.org/abs/2105.09316}{{\ttfamily 2105.09316}}].

\bibitem{Almheiri:2012rt}
A.~Almheiri, D.~Marolf, J.~Polchinski and J.~Sully, \emph{{Black Holes:
  Complementarity or Firewalls?}},
  \href{https://doi.org/10.1007/JHEP02(2013)062}{\emph{JHEP} {\bfseries 02}
  (2013) 062} [\href{https://arxiv.org/abs/1207.3123}{{\ttfamily 1207.3123}}].

\bibitem{Harlow:2013tf}
D.~Harlow and P.~Hayden, \emph{{Quantum Computation vs. Firewalls}},
  \href{https://doi.org/10.1007/JHEP06(2013)085}{\emph{JHEP} {\bfseries 06}
  (2013) 085} [\href{https://arxiv.org/abs/1301.4504}{{\ttfamily 1301.4504}}].

\bibitem{Akers:2022qdl}
C.~Akers, N.~Engelhardt, D.~Harlow, G.~Penington and S.~Vardhan, \emph{{The
  black hole interior from non-isometric codes and complexity}},
  \href{https://arxiv.org/abs/2207.06536}{{\ttfamily 2207.06536}}.

\bibitem{Engelhardt:2017aux}
N.~Engelhardt and A.~C. Wall, \emph{{Decoding the Apparent Horizon:
  Coarse-Grained Holographic Entropy}},
  \href{https://doi.org/10.1103/PhysRevLett.121.211301}{\emph{Phys. Rev. Lett.}
  {\bfseries 121} (2018) 211301}
  [\href{https://arxiv.org/abs/1706.02038}{{\ttfamily 1706.02038}}].

\bibitem{Engelhardt:2018kcs}
N.~Engelhardt and A.~C. Wall, \emph{{Coarse Graining Holographic Black Holes}},
  \href{https://doi.org/10.1007/JHEP05(2019)160}{\emph{JHEP} {\bfseries 05}
  (2019) 160} [\href{https://arxiv.org/abs/1806.01281}{{\ttfamily
  1806.01281}}].

\bibitem{Bousso:2019dxk}
R.~Bousso, V.~Chandrasekaran and A.~Shahbazi-Moghaddam, \emph{{From black hole
  entropy to energy-minimizing states in QFT}},
  \href{https://doi.org/10.1103/PhysRevD.101.046001}{\emph{Phys. Rev.}
  {\bfseries D101} (2020) 046001}
  [\href{https://arxiv.org/abs/1906.05299}{{\ttfamily 1906.05299}}].

\bibitem{Bousso:2015mna}
R.~Bousso, Z.~Fisher, S.~Leichenauer and A.~C. Wall, \emph{{Quantum focusing
  conjecture}}, \href{https://doi.org/10.1103/PhysRevD.93.064044}{\emph{Phys.
  Rev.} {\bfseries D93} (2016) 064044}
  [\href{https://arxiv.org/abs/1506.02669}{{\ttfamily 1506.02669}}].

\bibitem{Akers:2019lzs}
C.~Akers, N.~Engelhardt, G.~Penington and M.~Usatyuk, \emph{{Quantum Maximin
  Surfaces}}, \href{https://doi.org/10.1007/JHEP08(2020)140}{\emph{JHEP}
  {\bfseries 08} (2020) 140}
  [\href{https://arxiv.org/abs/1912.02799}{{\ttfamily 1912.02799}}].

\bibitem{Engelhardt:2021mue}
N.~Engelhardt, G.~Penington and A.~Shahbazi-Moghaddam, \emph{{A world without
  pythons would be so simple}},
  \href{https://doi.org/10.1088/1361-6382/ac2de5}{\emph{Class. Quant. Grav.}
  {\bfseries 38} (2021) 234001}
  [\href{https://arxiv.org/abs/2102.07774}{{\ttfamily 2102.07774}}].

\bibitem{Akers:2016aa}
C.~Akers, J.~Koeller, S.~Leichenauer and A.~Levine, \emph{Geometric constraints
  from subregion duality beyond the classical regime},
  \href{https://arxiv.org/abs/1610.08968}{{\ttfamily 1610.08968}}.

\bibitem{C:2013uza}
A.~C. Wall, \emph{{The Generalized Second Law implies a Quantum Singularity
  Theorem}}, \href{https://doi.org/10.1088/0264-9381/30/19/199501}{\emph{Class.
  Quant. Grav.} {\bfseries 30} (2013) 165003}
  [\href{https://arxiv.org/abs/1010.5513}{{\ttfamily 1010.5513}}].

\bibitem{Bousso:2022tdb}
R.~Bousso and A.~Shahbazi-Moghaddam, \emph{{Quantum Singularities}},
  \href{https://arxiv.org/abs/2206.07001}{{\ttfamily 2206.07001}}.

\bibitem{Bousso:2015eda}
R.~Bousso and N.~Engelhardt, \emph{{Generalized Second Law for Cosmology}},
  \href{https://doi.org/10.1103/PhysRevD.93.024025}{\emph{Phys. Rev.}
  {\bfseries D93} (2016) 024025}
  [\href{https://arxiv.org/abs/1510.02099}{{\ttfamily 1510.02099}}].

\bibitem{Bousso:2016aa}
R.~Bousso, Z.~Fisher, J.~Koeller, S.~Leichenauer and A.~C. Wall, \emph{Proof of
  the quantum null energy condition}, {\emph{Phys. Rev. D} {\bfseries 93}
  (2016) 024017} [\href{https://arxiv.org/abs/1509.02542}{{\ttfamily
  1509.02542}}].

\bibitem{Balakrishnan:2017bjg}
S.~Balakrishnan, T.~Faulkner, Z.~U. Khandker and H.~Wang, \emph{{A General
  Proof of the Quantum Null Energy Condition}},
  \href{https://doi.org/10.1007/JHEP09(2019)020}{\emph{JHEP} {\bfseries 09}
  (2019) 020} [\href{https://arxiv.org/abs/1706.09432}{{\ttfamily
  1706.09432}}].

\bibitem{Ceyhan:2018zfg}
F.~Ceyhan and T.~Faulkner, \emph{{Recovering the QNEC from the ANEC}},
  \href{https://arxiv.org/abs/1812.04683}{{\ttfamily 1812.04683}}.

\bibitem{Wall:2009wi}
A.~C. Wall, \emph{{Proving the Achronal Averaged Null Energy Condition from the
  Generalized Second Law}},
  \href{https://doi.org/10.1103/PhysRevD.81.024038}{\emph{Phys. Rev.}
  {\bfseries D81} (2010) 024038}
  [\href{https://arxiv.org/abs/0910.5751}{{\ttfamily 0910.5751}}].

\bibitem{Randall:1999vf}
L.~Randall and R.~Sundrum, \emph{{An Alternative to compactification}},
  \href{https://doi.org/10.1103/PhysRevLett.83.4690}{\emph{Phys. Rev. Lett.}
  {\bfseries 83} (1999) 4690}
  [\href{https://arxiv.org/abs/hep-th/9906064}{{\ttfamily hep-th/9906064}}].

\bibitem{Emparan:2006ni}
R.~Emparan, \emph{{Black hole entropy as entanglement entropy: A Holographic
  derivation}},
  \href{https://doi.org/10.1088/1126-6708/2006/06/012}{\emph{JHEP} {\bfseries
  06} (2006) 012} [\href{https://arxiv.org/abs/hep-th/0603081}{{\ttfamily
  hep-th/0603081}}].

\bibitem{Emparan:1999wa}
R.~Emparan, G.~T. Horowitz and R.~C. Myers, \emph{{Exact description of black
  holes on branes}},
  \href{https://doi.org/10.1088/1126-6708/2000/01/007}{\emph{JHEP} {\bfseries
  01} (2000) 007} [\href{https://arxiv.org/abs/hep-th/9911043}{{\ttfamily
  hep-th/9911043}}].

\bibitem{Verlinde:1999fy}
H.~L. Verlinde, \emph{{Holography and compactification}},
  \href{https://doi.org/10.1016/S0550-3213(00)00224-8}{\emph{Nucl. Phys.}
  {\bfseries B580} (2000) 264}
  [\href{https://arxiv.org/abs/hep-th/9906182}{{\ttfamily hep-th/9906182}}].

\bibitem{Gubser:1999vj}
S.~S. Gubser, \emph{{AdS / CFT and gravity}},
  \href{https://doi.org/10.1103/PhysRevD.63.084017}{\emph{Phys. Rev.}
  {\bfseries D63} (2001) 084017}
  [\href{https://arxiv.org/abs/hep-th/9912001}{{\ttfamily hep-th/9912001}}].

\bibitem{Myers:2013lva}
R.~C. Myers, R.~Pourhasan and M.~Smolkin, \emph{{On Spacetime Entanglement}},
  \href{https://doi.org/10.1007/JHEP06(2013)013}{\emph{JHEP} {\bfseries 06}
  (2013) 013} [\href{https://arxiv.org/abs/1304.2030}{{\ttfamily 1304.2030}}].

\bibitem{Chen:2020uac}
H.~Z. Chen, R.~C. Myers, D.~Neuenfeld, I.~A. Reyes and J.~Sandor,
  \emph{{Quantum Extremal Islands Made Easy, Part I: Entanglement on the
  Brane}}, \href{https://doi.org/10.1007/JHEP10(2020)166}{\emph{JHEP}
  {\bfseries 10} (2020) 166}
  [\href{https://arxiv.org/abs/2006.04851}{{\ttfamily 2006.04851}}].

\bibitem{Chen:2020hmv}
H.~Z. Chen, R.~C. Myers, D.~Neuenfeld, I.~A. Reyes and J.~Sandor,
  \emph{{Quantum Extremal Islands Made Easy, Part II: Black Holes on the
  Brane}}, \href{https://doi.org/10.1007/JHEP12(2020)025}{\emph{JHEP}
  {\bfseries 12} (2020) 025}
  [\href{https://arxiv.org/abs/2010.00018}{{\ttfamily 2010.00018}}].

\bibitem{Bousso:2020kmy}
R.~Bousso and E.~Wildenhain, \emph{{Gravity/ensemble duality}},
  \href{https://doi.org/10.1103/PhysRevD.102.066005}{\emph{Phys. Rev. D}
  {\bfseries 102} (2020) 066005}
  [\href{https://arxiv.org/abs/2006.16289}{{\ttfamily 2006.16289}}].

\bibitem{Karch:2000ct}
A.~Karch and L.~Randall, \emph{{Locally localized gravity}},
  \href{https://doi.org/10.1088/1126-6708/2001/05/008}{\emph{JHEP} {\bfseries
  05} (2001) 008} [\href{https://arxiv.org/abs/hep-th/0011156}{{\ttfamily
  hep-th/0011156}}].

\bibitem{Takayanagi:2011zk}
T.~Takayanagi, \emph{{Holographic Dual of BCFT}},
  \href{https://doi.org/10.1103/PhysRevLett.107.101602}{\emph{Phys. Rev. Lett.}
  {\bfseries 107} (2011) 101602}
  [\href{https://arxiv.org/abs/1105.5165}{{\ttfamily 1105.5165}}].

\bibitem{Fujita:2011fp}
M.~Fujita, T.~Takayanagi and E.~Tonni, \emph{{Aspects of AdS/BCFT}},
  \href{https://doi.org/10.1007/JHEP11(2011)043}{\emph{JHEP} {\bfseries 11}
  (2011) 043} [\href{https://arxiv.org/abs/1108.5152}{{\ttfamily 1108.5152}}].

\bibitem{Hubeny:2012wa}
V.~E. Hubeny and M.~Rangamani, \emph{{Causal Holographic Information}},
  \href{https://doi.org/10.1007/JHEP06(2012)114}{\emph{JHEP} {\bfseries 06}
  (2012) 114} [\href{https://arxiv.org/abs/1204.1698}{{\ttfamily 1204.1698}}].

\bibitem{Koeller:2016aa}
J.~Koeller and S.~Leichenauer, \emph{Holographic proof of the quantum null
  energy condition}, {\emph{Phys. Rev. D} {\bfseries 94} (2016) 024026}
  [\href{https://arxiv.org/abs/1512.06109}{{\ttfamily 1512.06109}}].

\bibitem{Hubeny:2007xt}
V.~E. Hubeny, M.~Rangamani and T.~Takayanagi, \emph{{A Covariant holographic
  entanglement entropy proposal}},
  \href{https://doi.org/10.1088/1126-6708/2007/07/062}{\emph{JHEP} {\bfseries
  07} (2007) 062} [\href{https://arxiv.org/abs/0705.0016}{{\ttfamily
  0705.0016}}].

\bibitem{Lewkowycz:2013nqa}
A.~Lewkowycz and J.~Maldacena, \emph{{Generalized gravitational entropy}},
  \href{https://doi.org/10.1007/JHEP08(2013)090}{\emph{JHEP} {\bfseries 08}
  (2013) 090} [\href{https://arxiv.org/abs/1304.4926}{{\ttfamily 1304.4926}}].

\bibitem{Fu:2017evt}
Z.~Fu, J.~Koeller and D.~Marolf, \emph{{The Quantum Null Energy Condition in
  Curved Space}}, \href{https://doi.org/10.1088/1361-6382/aa8f2c}{\emph{Class.
  Quant. Grav.} {\bfseries 34} (2017) 225012}
  [\href{https://arxiv.org/abs/1706.01572}{{\ttfamily 1706.01572}}].

\bibitem{Akers:2017ttv}
C.~Akers, V.~Chandrasekaran, S.~Leichenauer, A.~Levine and
  A.~Shahbazi~Moghaddam, \emph{{Quantum null energy condition, entanglement
  wedge nesting, and quantum focusing}},
  \href{https://doi.org/10.1103/PhysRevD.101.025011}{\emph{Phys. Rev. D}
  {\bfseries 101} (2020) 025011}
  [\href{https://arxiv.org/abs/1706.04183}{{\ttfamily 1706.04183}}].

\bibitem{Wall:2012uf}
A.~C. Wall, \emph{{Maximin Surfaces, and the Strong Subadditivity of the
  Covariant Holographic Entanglement Entropy}},
  \href{https://doi.org/10.1088/0264-9381/31/22/225007}{\emph{Class. Quant.
  Grav.} {\bfseries 31} (2014) 225007}
  [\href{https://arxiv.org/abs/1211.3494}{{\ttfamily 1211.3494}}].

\bibitem{Engelhardt:2019hmr}
N.~Engelhardt and S.~Fischetti, \emph{{Surface Theory: the Classical, the
  Quantum, and the Holographic}},
  \href{https://doi.org/10.1088/1361-6382/ab3bda}{\emph{Class. Quant. Grav.}
  {\bfseries 36} (2019) 205002}
  [\href{https://arxiv.org/abs/1904.08423}{{\ttfamily 1904.08423}}].

\bibitem{Sweerselliptic}
G.~Sweers, \emph{Strong positivity in{\$}{\$}c({$\backslash$}bar
  {$\backslash$}omega ){\$}{\$}for elliptic systemsfor elliptic systems},
  \href{https://doi.org/10.1007/BF02570833}{\emph{Mathematische Zeitschrift}
  {\bfseries 209} (1992) 251}.

\bibitem{Arvin:2023}
N.~Engelhardt, G.~Penington and A.~Shahbazi-Moghaddam, \emph{{Twice Upon a
  Time: Timelike-Separated Quantum Extremal Surfaces}}, {\emph{To Appear} }.

\bibitem{Engelhardt:2014aa}
N.~Engelhardt and A.~C. Wall, \emph{Quantum extremal surfaces: Holographic
  entanglement entropy beyond the classical regime},
  \href{https://arxiv.org/abs/1408.3203}{{\ttfamily 1408.3203}}.

\bibitem{renardy2006introduction}
M.~Renardy and R.~C. Rogers, \emph{An introduction to partial differential
  equations}, vol.~13. Springer Science \& Business Media, 2006.

\bibitem{Dvali:2000hr}
G.~R. Dvali, G.~Gabadadze and M.~Porrati, \emph{{4-D gravity on a brane in 5-D
  Minkowski space}},
  \href{https://doi.org/10.1016/S0370-2693(00)00669-9}{\emph{Phys. Lett. B}
  {\bfseries 485} (2000) 208}
  [\href{https://arxiv.org/abs/hep-th/0005016}{{\ttfamily hep-th/0005016}}].

\bibitem{Leichenauer:2018tnq}
S.~Leichenauer and A.~Levine, \emph{{Upper and Lower Bounds on the Integrated
  Null Energy in Gravity}},
  \href{https://doi.org/10.1007/JHEP01(2019)133}{\emph{JHEP} {\bfseries 01}
  (2019) 133} [\href{https://arxiv.org/abs/1808.09970}{{\ttfamily
  1808.09970}}].

\bibitem{Wall:2015raa}
A.~C. Wall, \emph{{A Second Law for Higher Curvature Gravity}},
  \href{https://doi.org/10.1142/S0218271815440149}{\emph{Int. J. Mod. Phys.}
  {\bfseries D24} (2015) 1544014}
  [\href{https://arxiv.org/abs/1504.08040}{{\ttfamily 1504.08040}}].

\bibitem{Leichenauer:2018obf}
S.~Leichenauer, A.~Levine and A.~Shahbazi-Moghaddam, \emph{{Energy density from
  second shape variations of the von Neumann entropy}},
  \href{https://doi.org/10.1103/PhysRevD.98.086013}{\emph{Phys. Rev. D}
  {\bfseries 98} (2018) 086013}
  [\href{https://arxiv.org/abs/1802.02584}{{\ttfamily 1802.02584}}].

\bibitem{Balakrishnan:2019gxl}
S.~Balakrishnan, V.~Chandrasekaran, T.~Faulkner, A.~Levine and
  A.~Shahbazi-Moghaddam, \emph{{Entropy Variations and Light Ray Operators from
  Replica Defects}},  \href{https://arxiv.org/abs/1906.08274}{{\ttfamily
  1906.08274}}.

\end{thebibliography}\endgroup

\end{document}